# RSSI-CSI Measurement and Variation Mitigation with Commodity WiFi Device

Bo Wei, Hang Song, Jiro Katto, and Takamaro Kikkawa

*Abstract*—Owing to the plentiful information released by the commodity devices, WiFi signals have been widely studied for various wireless sensing applications. In many works, both received signal strength indicator (RSSI) and the channel state information (CSI) are utilized as the key factors for precise sensing. However, the calculation and relationship between RSSI and CSI is not explained in detail. Furthermore, there are few works focusing on the measurement variation of the WiFi signal which impacts the sensing results. In this paper, the relationship between RSSI and CSI is studied in detail and the measurement variation of amplitude and phase information is investigated by extensive experiments. In the experiments, transmitter and receiver are directly connected by power divider and RF cables and the signal transmission is quantitatively controlled by RF attenuators. By changing the intensity of attenuation, the measurement of RSSI and CSI is carried out under different conditions. From the results, it is found that in order to get a reliable measurement of the signal amplitude and phase by commodity WiFi, the attenuation of the channels should not exceed 60 dB. Meanwhile, the difference between two channels should be lower than 10 dB. An active control mechanism is suggested to ensure the measurement stability. The findings and criteria of this work is promising to facilitate more precise sensing technologies with WiFi signal.

## I. Introduction

TAKING advantages of the ubiquitous wireless signals for sensing has attracted interest from wide range of researchers [1], [2]. Especially, the WiFi signal is considered to be promising for precise wireless sensing, owing to the fruitful information released by the commodity devices [3]. Among the released information, the received signal strength indicator (RSSI) and channel state information (CSI) are extremely essential. RSSI is a kind of preliminary data which represents the aggregative power of the entire signal bandwidth. However, it is a coarse information, which may limit the sensing accuracy if only RSSI is exploited for sensing. On the other hand, CSI provides the fine-grained channel responses of different subcarriers in physical layer, which can enable more precise sensing ability.

Recently, WiFi signal has been applied to various internet of things (IoT) applications such as motion detection [4]-[7], positioning and localization [8]-[10], security surveillance [11]-[14], health monitoring [15]-[19], and material identification [20], [21]. During the development of different applications, both RSSI and CSI data of the WiFi signals were utilized. Some works utilized only one kind of data such as RSSI [4, 11], only CSI phase [5, 16] or CSI amplitude [8, 9, 12, 14, 15]. Others combined different data categories [6, 7, 10, 13, 17-21]. However, it is found that the scales of the measured CSI amplitude in various works are quite different. This may limit the application of these works to other scenarios. Therefore, it is necessary to figure out a calculation method for getting comparable measurement results.

As for the phase of CSI, there have been many studies focusing on how to extract the phase information [22]-[27]. Linear transformation is a preliminary method which subtracts the raw phase with a linear slope [22], [23]. However, this method may add other errors. Some works proposed linear fitting method to subtract the linear slope from the raw phase [24], [25]. But the problem of introducing additional errors remains. Many other works utilized a differential method to extract the phase [16]-[19], [26]. In this method, the phase difference between two receiving channels on the same chip is taken and the phase distortions can be eliminated. Therefore, the differential approach is considered to be suitable for various applications. Although the extraction of phase information has been comprehensively studied, the measurement variation has not been investigated yet when the receiving powers at different channels are diverse.

From literature review, numerous methods and algorithms have been proposed to improve the sensing accuracy for different purposes [4]-[21]. However, there are few works dealing with the measurement variation of the WiFi signal. If the measurement itself has large variations and the measured data deviate severely from the true values, even advanced algorithms may be unable to generate precise sensing result. Therefore, it is essential to analyze the measurement variation and find solutions to obtain stable measurement results.

In this paper, the measurement of RSSI and CSI and the relationship between RSSI and CSI is studied to figure out a calculation method for obtaining comparable measurement results. Furthermore, the measurement variation of amplitude and phase information is investigated by extensive experiments. In order to quantitatively control the loss of the communication channels, RF cables and attenuators are utilized to directly connect the transmitter and receiver, which are the Intel 5300 NICs. The measurement of RSSI and CSI and the variations are

Bo Wei and Jiro Katto are with the Department of Computer Science and Communication Engineering, Waseda University, Tokyo 169-8555, Japan.

Hang Song is with School of Engineering, The University of Tokyo, Tokyo 113-8654, Japan.

Takamaro Kikkawa is with the Research Institute for Nanodevice and Bio Systems, Hiroshima University, Hiroshima 739-8527, Japan.

analyzed under different conditions, where the intensity of attenuation of three channels are configured as various values. Through the experiments, it is found that the accuracy of the measurement will decrease and the variation will increase dramatically when the attenuation is larger than 60 dB. Besides, if the difference between two channels is over 30 dB, the phase information cannot be measured. Therefore, in order to get a reliable measurement of the amplitude and phase information by commodity WiFi, the attenuation of the channel should not exceed 60 dB. Meanwhile, the difference between two antennas should be lower than 10 dB. Based on the findings, an active control mechanism is suggested to ensure the measurement stability and mitigate variation. To the best of our knowledge, this is the first work which focuses on the measurement variation with commodity WiFi device. This work provides criteria which is promising to empower future design of precise sensing technologies with WiFi signal.

The rest of this paper is organized as follows. Section II gives a brief review of the works which utilized various WiFi signal data for different applications. In Section III, the calculation and the relationship between RSSI and CSI are detailed. Section IV depicts the experiment setup for measuring the RSSI and CSI. Section V presents the measurement results under different conditions and the measurement variation is analyzed. Finally, the conclusion is made in Section VI.

## II. RELATED WORK

By exploiting the RSSI and CSI data of the WiFi signals, a lot of systems and algorithms have been developed for different IoT applications.

In motion detection, Gu *et al.* [4] developed an online fingerprint-based activity recognition system to distinguish six different activities, where only RSSI data were utilized. The fusion algorithm was also designed to improve the accuracy. Since it was found that RSSI is difficult to deal with the multipath effect, MoSense was designed where the phase information of CSI was utilized to overcome this shortage [5]. Algorithms were also developed to characterize stationary state and exclude noisy channels. Duan *et al.* [6] proposed a neural network-based method for action recognition, where both the amplitude and phase information of CSI were utilized. In [7], Xiao *et al.* incorporated the generative adversarial network (GAN) technique to enhance the recognition performance.

In localization, Wang *et al.* [8] proposed LIFS to realize device-free localization, where the amplitude information of CSI was utilized. A pre-processing scheme was proposed to identify clean subcarriers. In [9], Liu *et al.* proposed C-Map which also utilized CSI amplitude for fingerprint localization. In C-Map, the dynamic denoising and feature enhancement methods were proposed to reduce the position error. In [10], Zhang *et al.* proposed a position method, in which the ratio of the amplitude and phase information of CSI was optimized. LSTM neural network was utilized to train the data sets with different proportions.

In security, Cheng *et al.* proposed NiFi to identify different users where only the RSSI was utilized [11]. By analyzing the dynamics of the signal sequences, NiFi was able to recognize undesired users. Lin *et al.* developed PetFree to reduce the influence of pets on the intrusion detection [12]. PetFree used the amplitude information of CSI and proposed the effective interference height model to differentiate humans and pets. Shi *et al.* designed a CNN-based architecture for user authentication where both the amplitude and phase information of CSI were utilized [13]. Cao *et al.* proposed LW-WiID, a lightweight deep learning identification model to compress the model size and improve the training speed [14]. In this model, the amplitude of CSI was utilized.

In health monitoring, Liu *et al.* utilized the amplitude of CSI to estimate the breathing rate, heart rate and posture during sleep [15]. Zhang *et al.* developed BreathTrack which employed the phase information of CSI to track the breath status [16]. Zeng *et al.* designed a real-time respiration detection system by exploiting both the amplitude and phase of CSI, which could achieve full location coverage and eliminate blind spots [17]. Wang *et al.* proposed a contactless fall detection utilizing both amplitude and phase information CSI [18]. Yu *et al.* developed WiFi-sleep which can extract information of the respiration and body movement [19]. In WiFi-sleep, both amplitude and phase of CSI were utilized and advanced signal processing methods with deep learning were proposed.

In material identification, Feng *et al.* proposed WiMi to identify different materials, where both the amplitude and phase information of CSI were utilized [20]. In our former work, WiEps was developed which can quantitatively measure the dielectric properties of material, by combining both RSSI and CSI data [21].

## III. RSSI AND CSI CALCULATION AND THEIR RELATIONSHIP

As reviewed in Section II, plenty of works utilized the RSSI and CSI information to develop wireless sensing technologies. However, the calculated amplitude of the CSI is very different in scale. Since the measured CSI data is a kind of relative values, each work may have their own calculation approaches. Based on the calculated values, the developed methods and algorithms achieved good results. But the difference in CSI calculation may limit the application of these works to other scenarios. In this section, a calculation method is provided based on the operation of the WiFi chip circuits.

Although the detailed structure of the WiFi chips is not available, the general structure can be illustrated as Fig. 1. On the receiver side of the WiFi communication, the signals are firstly captured by the antennas and then amplified by amplifiers. It should be noted that, there is an adaptive amplification mechanism, automatic gain control (AGC), to control the signal power before analog-digital conversion (ADC). The AGC can ensure the accuracy of the measurement. Meanwhile, it can also ensure that signal level does not exceed the safety limitation of the circuits in the chip. After the amplification, the signals are divided into in-phase (I) path and quadrature (Q) path. And the I and Q are digitized separately by ADC. Then, the I and Q data are merged to generate the CSI. Meanwhile, the power of the signal is also digitized to generate RSSI. There are three ports to receive signals on the chip and the amplification of the three channels are the same. The RSSI and CSI are digitized separately for three channels.

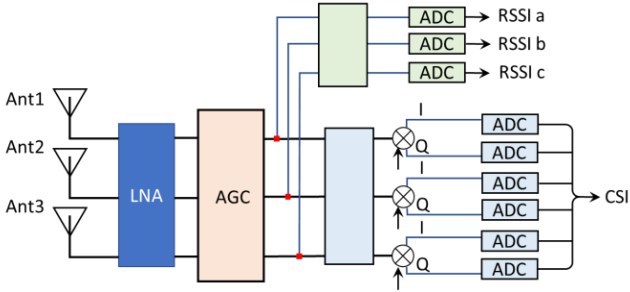

**Fig. 1.** Schematic illustration of the receiver side in WiFi communication.

For the measured RSSI, the values are in decibel (dB) and relative to the AGC and other amplifiers. Therefore, in order to obtain the power of the received signal, the amplification factors need to be removed. For each channel, the received power in dBm is calculated as follows.

$$P_a(\text{dBm}) = RSSI_a - AGC - C \quad (1)$$

$$P_b(\text{dBm}) = RSSI_b - AGC - C \quad (2)$$

$$P_c(\text{dBm}) = RSSI_c - AGC - C \quad (3)$$

where AGC is adaptive amplification factor included in the CSI information package. C is the fixed factor which is the total effect of other amplifiers and the loss by other modules. It is given in the CSI tool [3]. Then, the total received power of the three channels can be calculated as:

$$\begin{aligned} P_{total}(\text{dBm}) &= 10 \log_{10}\left(10^{\frac{P_a}{10}} + 10^{\frac{P_b}{10}} + 10^{\frac{P_c}{10}}\right) \\ &= 10 \log_{10}\left[\left(10^{\frac{RSSI_a}{10}} + 10^{\frac{RSSI_b}{10}} + 10^{\frac{RSSI_c}{10}}\right)\left(10^{\frac{-AGC-C}{10}}\right)\right] \\ &= 10 \log_{10}\left(10^{\frac{RSSI_a}{10}} + 10^{\frac{RSSI_b}{10}} + 10^{\frac{RSSI_c}{10}}\right) - AGC - C \end{aligned} \quad (4)$$

Channel state information reveals the channel response of the communication at each subcarrier frequency $f$ and different timing $t$. It is generally expressed as the superposition of multipath effect:

$$\begin{aligned} H(f,t) &= \sum_{n=1}^{N} A_n(f,t) \cdot e^{j\theta_n(f,t)} \\ &= A(f,t) \cdot e^{j\theta(f,t)} \end{aligned} \quad (5)$$

where $A_n(f,t)$ is the amplitude attenuation of $n$th signal path and $\theta_n(f,t)$ is the phase change. For the measured CSI, the values are provided in the complex and linear form.

$$H_{\text{mea}}(f,t) = \alpha(f,t) + j\beta(f,t) \quad (6)$$

As shown in Fig. 1, the digitized CSI data are affected by the AGC and other modules. Therefore, the nominal amplitude of the CSI, which is $\sqrt{\alpha^2 + \beta^2}$, does not exhibit the actual channel attenuation. In some works, the nominal amplitude is used directly and others transform it into logarithmic form. Using the relative values, the developed sensing algorithms work well in the test scenarios. However, when applying the algorithm to other scenarios where the environment and the signal are very different, the relative CSI amplitude may not correctly reflect the channel attenuation status. This will impair the sensing accuracy. Therefore, the actual amplitude of CSI is necessary to indicate the real channel condition.

The calculation of the amplitude of CSI cannot be achieved straightforwardly because the measurement of CSI is carried out by other modules and ADCs. The ratio between the real value and the measured value is unknown. Fortunately, since the RSSI reflects the aggregative power of all subcarriers, the amplitude of CSI can be estimated exploiting RSSI data. Theoretically, the ratio of power of three channels is the same as that of the sum of CSI amplitude squared:

$$P_a : P_b : P_c = \sum_{f=f_0}^{f_K} A_1^2(f) : \sum_{f=f_0}^{f_K} A_2^2(f) : \sum_{f=f_0}^{f_K} A_3^2(f) \quad (7)$$

However, it should be noted that both the RSSI and CSI are digitized, and they are measured in logarithmic and linear form, respectively. Therefore, the relationship between RSSI and CSI can be expressed as:

$$RSSI_b - RSSI_a = 10 \log_{10} \frac{\sum_{f=f_0}^{f_K}[\alpha_2^2(f)+\beta_2^2(f)]}{\sum_{f=f_0}^{f_K}[\alpha_1^2(f)+\beta_1^2(f)]} \quad (8)$$

$$RSSI_c - RSSI_b = 10 \log_{10} \frac{\sum_{f=f_0}^{f_K}[\alpha_3^2(f)+\beta_3^2(f)]}{\sum_{f=f_0}^{f_K}[\alpha_2^2(f)+\beta_2^2(f)]} \quad (9)$$

$$RSSI_a - RSSI_c = 10 \log_{10} \frac{\sum_{f=f_0}^{f_K}[\alpha_1^2(f)+\beta_1^2(f)]}{\sum_{f=f_0}^{f_K}[\alpha_3^2(f)+\beta_3^2(f)]} \quad (10)$$

Considering the quantum errors during the ADC, there will be slight difference on two sides of the equation in (8)~(10). Since the RSSI is digitized in logarithmic form, the resolution is lower than the CSI which is in linear form. To restore the CSI amplitude, a scale factor is calculated in terms of the total received power:

$$\rho = \frac{10^{\frac{P_{total}}{10}}}{\sum_{i=1}^{3}\sum_{f=f_0}^{f_K}[\alpha_i^2(f)+\beta_i^2(f)]} \quad (11)$$

Multiplying the nominal amplitude of CSI by $\rho$, the amplitude of the channel response can be obtained in dBm:

$$A(\text{dBm}) = 10 \log_{10}\{\rho * [\alpha^2 + \beta^2]\} \quad (12)$$

Processing with (12), the calculated amplitude of CSI reflects the actual attenuation status of the communication channel.

For the phase of CSI, the raw information $\arctan(\beta/\alpha)$ cannot reveal the real phase change of the channel since the transmitter and receiver are not synchronized. Factors such as carrier frequency offset (CFO), packet detection delay (PDD), and sampling frequency offset (SFO) will cause the raw phase change randomly. The raw phase can be expressed as:

$$\arctan\left(\frac{\beta(f,t)}{\alpha(f,t)}\right) = \theta(f,t) + \theta_{CFO}(f,t) \\ + \theta_{PDD}(f,t) + \theta_{SFO}(f,t) + \delta_i \quad (13)$$

where $\theta$ is the actual phase change of the channel. $\theta_{CFO}$, $\theta_{PDD}$, $\theta_{SFO}$ are the time-variant phase offsets caused by CFO, SFO, and PDD, respectively. $\delta_i$ is a constant phase offset in $i$th channel caused by the phase-locked-loop (PLL) locking point. Several methods have been proposed to extract the phase information from the measured CSI data [22]-[27]. Among those, the differential method is widely adopted for its simplicity and efficacy. This method is based on the finding that the factors which cause random phase offset are the same among the three channels on the same WiFi chip. Therefore, by differentiating the raw phases of two channels, the random factors can be eliminated and the phase difference becomes stable. By setting a reference channel, the phase of CSI on other channels can be estimated.

## IV. EXPERIMENT SETUP

To evaluate the calculation of CSI and the relationship between measured RSSI and CSI, experiments in the real world were conducted. The experiment setup is shown in Fig. 2. Two Intel 5300 NICs were utilized as the transmitter and receiver, respectively. On the transmitter side, one port was exploited for transmitting signal. The other two ports were connected to two 50 Ω terminators to avoid reflections. In order to quantitatively control the condition of the communication channel, the transmitter and receiver were connected by cables, and a 4-way power divider and attenuators were inserted in the propagation paths. The function of the power divider was to split the input power into several equal shares. The input was connected to the transmitter and 3 outputs were connected to the three ports of the receiver. The unused output was connected to a 50 Ω terminator. Between the power divider and the receiver, attenuators were inserted separately into the three channels.

During the experiment, the attenuators were independently adjusted to mimic various path loss conditions. For each test, the attenuators were set to different values and other components were kept the same. Then, the CSI data were captured and saved on the receiver side. After all tests were completed, the CSI data were extracted and processed using the methods depicted in Section II. Subsequently, the RSSI as well as the amplitude and phase of CSI were analyzed.

## V. MEASUREMENT RESULTS AND VARIATION ANALYSIS

### A. RSSI and CSI Calculation Results

Two measurement experiments were firstly conducted to verify the calculation method. The attenuation intensities of the three channels were set as 66 dB, 66 dB, 66 dB and 33 dB, 30 dB, 36 dB, respectively.

The measured values of RSSI are shown in Table I. For both tests, it can be observed that the measured RSSIs almost reflect the attenuation extents of different channels. In test 1, the difference between port 1 and 2 is 1 dB, while the set attenuations are the same. In test 2, the difference between port 2 and 3 is 8 dB while the set difference was 6 dB. Although the attenuation setting in two tests had a difference of 30 dB, the nominal values of the measured RSSIs are within the same range. This indicates that the nominal RSSI only reflects the relative difference among channels but cannot reflect the attenuation change in the channel. By considering the effect of AGC and other modules, the actual RSSI is calculated in dBm. It can be observed that the calculated RSSI exhibits both the channel attenuation change and the difference between channels.

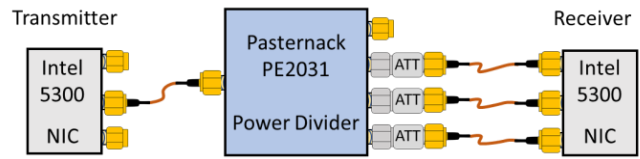

Fig. 2. Schematic of experiment setup for evaluation.

TABLE I
MEASURED AND CALCULATED VALUES OF RSSI

|  | Test 1 | | | Test 2 | | |
| --- | --- | --- | --- | --- | --- | --- |
|  | Port 1 | Port 2 | Port 3 | Port 1 | Port 2 | Port 3 |
| Attenuation (dB) | 66 | 66 | 66 | 33 | 30 | 36 |
| Measured RSSI | 37 | 38 | 37 | 36 | 39 | 31 |
| AGC | 62 | 62 | 62 | 28 | 28 | 28 |
| Calculated RSSI (dBm) | -69 | -68 | -69 | -36 | -33 | -41 |

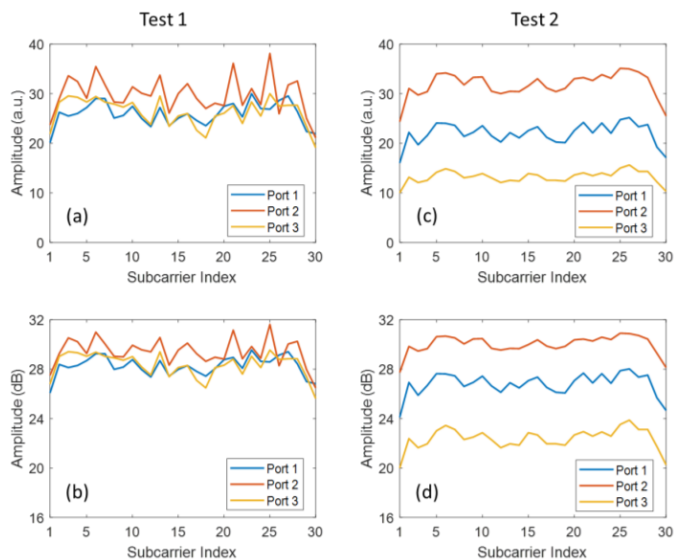

Fig. 3. Raw amplitudes of CSI at different subcarriers. (a) Linear form and (b) logarithmic form of Test 1. (c) Linear form and (d) logarithmic form of Test 2.

The raw amplitudes of the CSI data which are the modulus of the recorded complex values are shown in Fig. 3. From the results, it can be observed that the amplitudes are almost the same in Fig. 3(a) and those in Fig. 3(c) have large deviation. This observation is consistent with the experiment setting and indicates that the relative relationship between different channels is demonstrated in the raw amplitude. On the other hand, comparing Fig. 3(a) and (c), the intensities of port 2 in the two tests are with the same level around 30, although the actual attenuation difference between the two cases is 36 dB. The amplitudes transformed into logarithmic form are shown in Fig. 3(b) and (d). The values in dB are also consistent with the attenuation settings but they cannot show the loss change in the channel, either. Therefore, it can be concluded that the raw amplitude of CSI is not appropriate to represent the absolute channel loss.

Table II shows the power difference between channels evaluated by RSSI and CSI. The aggregative power of CSI is calculated using equations depicted in (8)~(10). The values by RSSI are integers while those by CSI have decimal numbers. This is because the RSSI is digitized directly in logarithmic form and the CSI are recorded in linear form. Therefore, the data precision is higher in CSI. As can be seen, the power difference calculated by CSI is almost the same as that by RSSI. This observation confirms that the ratio of amplitude among different channels can be also expressed by CSI. Thus, exploiting the total power calculated with RSSI, the absolute change of CSI amplitude can be estimated.

The amplitudes of CSI calculated with (10)~(11) are shown in Fig. 4. The unit of the amplitude is in dBm. It can be observed that both the ratio among channels and the loss change in the channel are reflected in the amplitude, demonstrating the correctness and efficacy of the calculation method of CSI proposed in Section II.

TABLE II
POWER DIFFERENCE BY CSI AND RSSI

| Difference | Test 1 | | | Test 2 | | |
|---|---|---|---|---|---|---|
| | 2/1 | 3/2 | 1/3 | 2/1 | 3/2 | 1/3 |
| Set value | 0 | 0 | 0 | 3 | -6 | 3 |
| By RSSI | 1 | -1 | 0 | 3 | -8 | 5 |
| By CSI | 1.23 | -1.10 | -0.12 | 3.19 | -7.60 | 4.41 |

Unit: dB

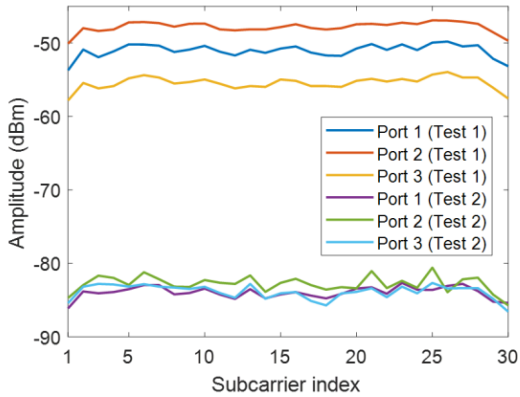

**Fig. 4.** Amplitudes of CSI at different subcarriers by the proposed method.

The phases of CSI are calculated by differential method and the phase of the 15$^{th}$ subcarrier along receiving time is shown in Fig. 5. It can be observed that the raw phase of CSI randomly fluctuates all the time although the channel is static during the test. On the contrary, the phase difference between two ports is stable along the time, demonstrating the efficacy of the differential method for extracting the phase information of CSI. Comparing the phase in Test 1 and Test 2, it can be noted that measurement variation is larger in Test 1. It is caused by the weak received signal strength. The measurement variation is comprehensively investigated in the subsequent parts by extensive experiments under different conditions.

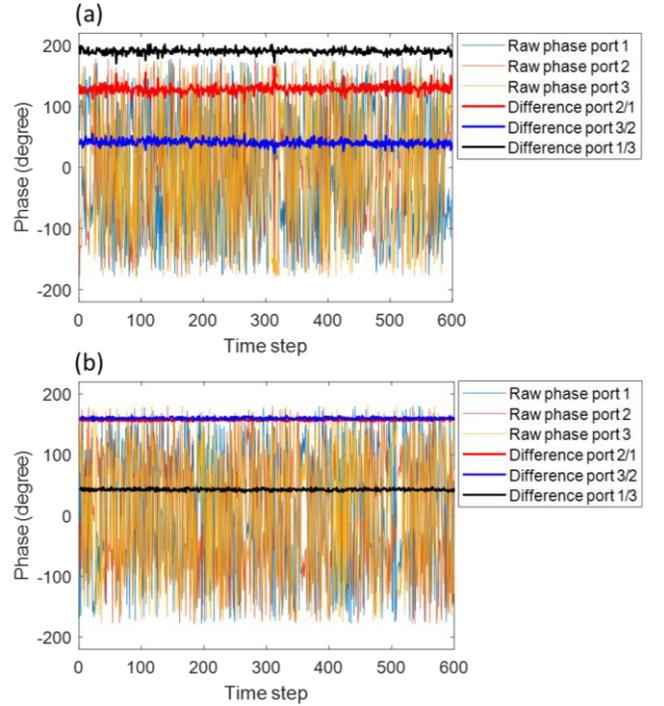

**Fig. 5.** Phase of CSI at different time step. (a) Raw phase and phase difference of Test 1. (b) Raw phase and phase difference of Test 2.

### B. Measurement Variation under Different Conditions

Two groups of experiments were conducted by configuring different attenuation extents of the channels. In the first group, the attenuations of the three channels were set to be the same but with different intensities. The settings of the attenuators are shown in Table III. In the second group, the three channels were set with different arbitrary loss extents. The settings of the attenuators are shown in Table IV.

TABLE III
ATTENUATION SETTING IN GROUP ONE

| Test Order | Attenuation (dB) | | | | | | |
|---|---|---|---|---|---|---|---|
| 1 to 7 | 16 | 19 | 20 | 26 | 30 | 36 | 40 |
| 8 to 14 | 46 | 50 | 56 | 60 | 66 | 70 | 80 |

TABLE IV
ATTENUATION SETTING IN GROUP TWO

| Test Order | Attenuation (dB) | | | Test Order | Attenuation (dB) | | |
|---|---|---|---|---|---|---|---|
| | Port 1 | Port 2 | Port 3 | | Port 1 | Port 2 | Port 3 |
| 1 | 23 | 50 | 50 | 10 | 56 | 70 | 26 |
| 2 | 30 | 80 | 80 | 11 | 23 | 26 | 20 |
| 3 | 50 | 50 | 30 | 12 | 40 | 43 | 46 |
| 4 | 26 | 56 | 40 | 13 | 33 | 30 | 36 |
| 5 | 66 | 20 | 36 | 14 | 50 | 56 | 53 |
| 6 | 60 | 56 | 66 | 15 | 40 | 20 | 30 |
| 7 | 53 | 50 | 63 | 16 | 30 | 40 | 50 |
| 8 | 56 | 40 | 36 | 17 | 50 | 60 | 40 |
| 9 | 23 | 60 | 30 | | | | |

### C. Signal Strength Measurement Results

Figure 6 shows the measured RSSI power level of the received signal under the conditions in group one, when the attenuation was set the same for three channels. The black line in Fig. 6(a) depicts the theoretical relationship between the attenuation and the power level. It can be observed that the RSSI power level agrees well with the theoretical line except for the cases when the attenuation is larger than 66 dB or smaller than 20 dB. The deviation of the RSSI is shown in Fig. 6(b). The absolute value of deviation is less than 2 dB when the attenuation is within [20 dB, 66 dB]. Inspecting the AGC of the cases outside this range, it is found that the AGC values are fixed as 26 when the attenuation is 16 dB and 19 dB. While those are fixed as 63 when the attenuation is 70 dB and 80 dB. It implied that the minimum amplification ability of AGC is 26 dB. When the input power is larger, the circuits cannot attenuate the signal to the proper power level for ADC, thus causes the inaccuracy in measured RSSI power level. On the other hand, it also implied that the maximum amplification ability of AGC is 63 dB. If the path loss is too severe, the RSSI cannot measure the power level correctly.

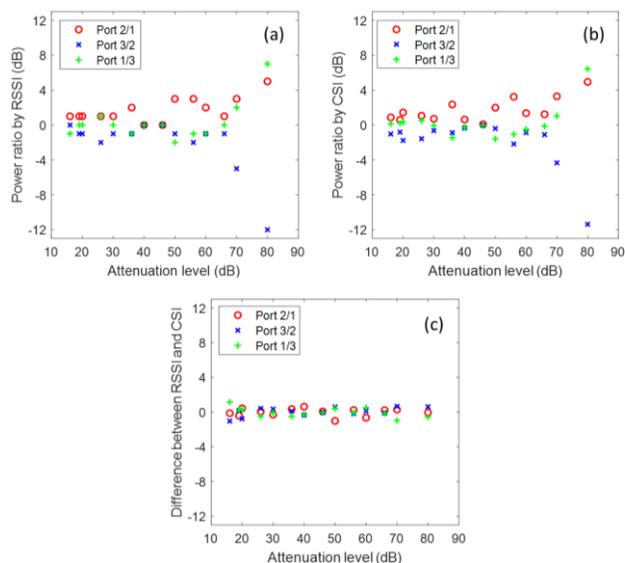

**Fig. 7.** Power ratio between channels when the attenuation was set the same among three channels. (a) Power ratio calculated by RSSI. (b) Power ratio calculated by CSI. (c) Difference of calculated power ratios by RSSI and CSI.

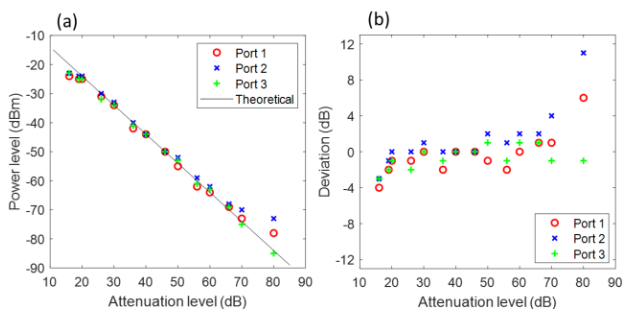

**Fig. 6.** Measured signal power levels when the attenuation was set the same among three channels. (a) RSSI power level at different attenuation conditions. (b) Deviation of RSSI from the theoretical value.

The power ratios between channels calculated by RSSI and CSI are shown in Figure 7. Since the attenuation was set the same in this group, the power ratio should be 0 dB theoretically. It can be observed in Fig. 7(a), the absolute deviation of power ratio dramatically increases when the attenuation is larger than 66 dB. This is consistent with the observation about the power level. However, the deviation is 3 dB when the attenuations are 50 dB and 56 dB. This is considered to be caused by the measurement error of the chip circuits. If the measured value is larger than the true value in one channel and smaller in another channel, then the difference between two channels will be larger. Similar observation can also be seen in the power ratio calculated by CSI as shown in Fig. 7(b). Comparing the results between RSSI and CSI, it is found that the deviation values are very similar even when the attenuations are 70 dB and 80 dB. The difference of the calculated power ratio is shown in Fig. 7(c) and the absolute values are almost less than 1 dB. This indicates that the relationship between RSSI and CSI expressed in (8)~(10) is satisfied regardless of the received signal power strength and the measurement accuracy by the chip circuits.

The measured RSSI power level of the received signal under the conditions in group two is illustrated in Fig. 8. The attenuations of three channels were different in this group. It can be observed that the measured RSSI power levels also show good agreement with the theoretical values. The deviation of the RSSI is shown in Fig. 8(b). In group two, the deviation is less than 2 dB when the attenuation is smaller than 56 dB, expect in one case of 36 dB attenuation. The attenuations in this case were set as 66 dB, 20 dB, 36 dB, respectively. The strongest signal was from port 2. The readouts of RSSI are 3, 46, 24, respectively and the AGC is 26. This implied that the chip circuits tend to preferentially ensure the power of the strongest channel does not exceed the safety threshold and the AGC is mainly affected by the strongest channel. When the attenuation difference between channels is relatively large, the accuracy of the measured power level will decrease in the channels with weak signal power.

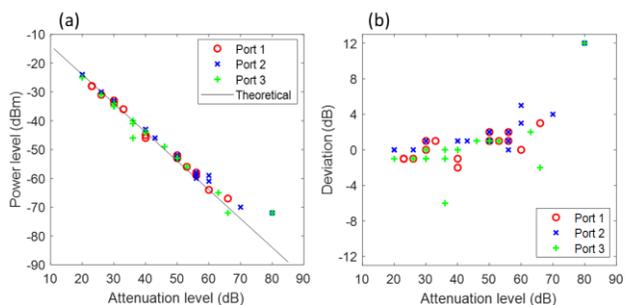

**Fig. 8.** Measured signal power levels when the attenuation was set differently among three channels. (a) RSSI power level at different attenuation conditions. (b) Deviation of RSSI from the theoretical value.

Figure 9 shows the power ratios between channels in group two calculated by RSSI and CSI. It can be observed in Fig. 9(a) that when the difference between channels increases, the accuracy of power ratio measured by RSSI decreases. The deviation of measured power ratio from the theoretical values is shown in Fig. 9(b). It indicates that the absolute value of deviation is less than 2 dB when the attenuation difference is less than 10 dB. This observation is consistent with the finding that large attenuation difference will impair the measurement accuracy. Power ratio calculated by CSI and the deviation are shown in Fig. 9(c) and (d), respectively. Similar behavior is observed that when the attenuation difference is larger than 10 dB, the deviation increases dramatically. The difference of the calculated power ratio by RSSI and CSI is shown in Fig. 9(e). It can be observed that the disagreement is less than 1 dB when the attenuation difference is less than 10 dB, except for one case in 0 dB. The attenuations in this special case were set as 30 dB, 80 dB, 80 dB, respectively. Since the strongest signal channel had 50 dB power difference with the other two, the measurement accuracy drops severely. These results indicates that the relationship between RSSI and CSI expressed in (8)~(10) is still satisfied if the attenuation difference between channels is less than 10 dB.

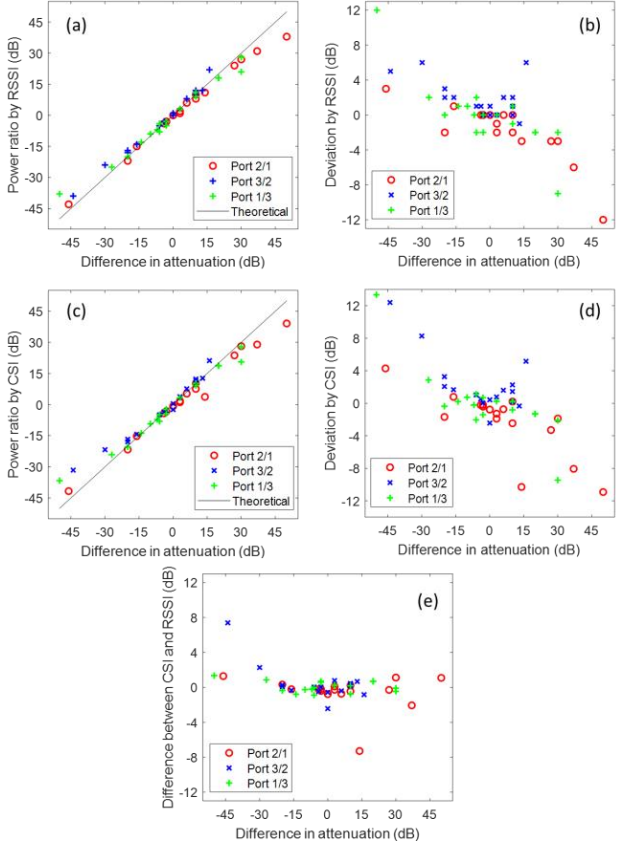

**Fig. 9.** Power ratio between channels when the attenuation was set differently among three channels. (a) Power ratio calculated by RSSI and (b) Deviation from theoretical value. (c) Power ratio calculated by CSI and (d) Deviation from theoretical value. (e) Difference of calculated power ratios by RSSI and CSI.

From the results, it is found that in order to ensure the measurement accuracy for the signal power level, the attenuation of the propagation channel should be less than 60 dB as well as larger than 20 dB. Meanwhile, the difference in received signal power between channels should be less than 10 dB. Meeting these criteria, the power ratio depicted by RSSI shows good agreement of that calculated by CSI.

*D. Measurement Variation in Amplitude and Phase of CSI*

Using the calculation method in (11)~(12), the amplitude of the CSI is obtained. Meanwhile, the phase information of CSI is extracted by using the differential method. Figure 10 shows the amplitude and phase information of CSI at $15^{th}$ subcarrier in group one as an example. In the figure, the points in one cluster are the results at different time steps. Three clusters of the same color represent the results for one case. It can be observed that the measurement variation increases when the attenuation increases. The standard deviation (STD) of the amplitude and phase of CSI is shown in Fig. 11. When the attenuation is lower than 60 dB, the amplitude STD is less than 0.5 dB. On the contrary, the amplitude STD increases significantly when the attenuation is larger than 60 dB. In terms of phase, the STD is less than $2°$ when the attenuation is lower than 50 dB. The STD increase to about $3°$ when the attenuation is 60 dB. After that, the STD increases radically when the attenuation is larger than 60 dB.

From the results, it can be inferred that to get a stable measurement of amplitude and phase of CSI, the loss of the channel should be less than 60 dB.

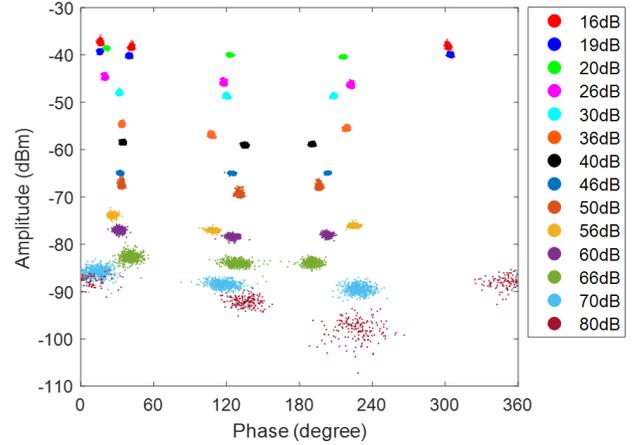

**Fig. 10.** Amplitude and phase of CSI at the $15^{th}$ subcarrier under different conditions where attenuation was set the same among three channels.

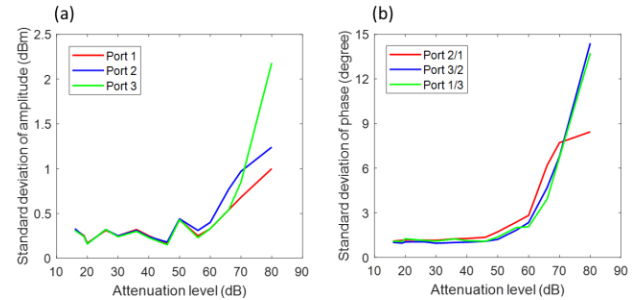

**Fig. 11.** Standard deviation of (a) amplitude and (b) phase of the CSI.

In the $2^{nd}$, $5^{th}$, $9^{th}$, and $10^{th}$ tests of group two, the measured amplitude and phase of CSI at $15^{th}$ subcarrier are shown in Fig. 12. The red circle represents the amplitude of port 1 and the phase difference between port 2 and port 1. The blue cross presents the amplitude of port 2 and the phase difference between port 3 and port 2. The green plus sign the amplitude of port 3 and the phase difference between port 1 and port 3. It can be observed that there are phases which cannot be measured in these cases. This is caused by the large difference in signal strength between channels. By inspecting the obtained raw CSI, it is found that there were zero values across all subcarriers in the channel with the largest attenuation. From these observations, it suggests that the phase information cannot be extracted if the difference of signal strength between two channels is larger than 30 dB.

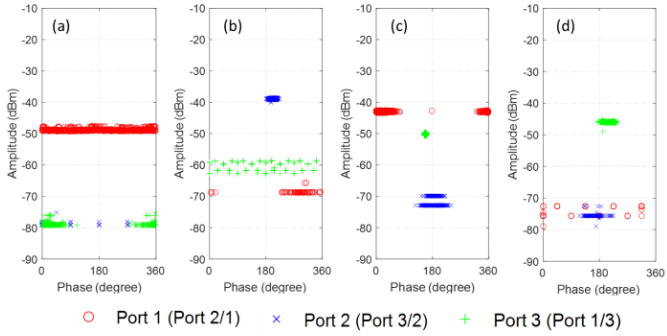

**Fig. 12.** Amplitude and phase of CSI at the 15th subcarrier under different conditions. Attenuations of three channels were set as (a) 30 dB, 80 dB, 80 dB; (b) 66 dB, 20 dB, 36 dB; (c) 23 dB, 60 dB, 30 dB; (d) 56 dB, 70 dB, 26 dB.

The standard deviation (STD) of amplitude and phase of CSI in other tests of group two are shown in Table V. Based on the maximum attenuation difference, the results are divided into three sets. In Set 1, the maximum attenuation difference is around 30 dB and the phase STD is almost from 8° to 16°. The phase STD is only 3° for port 1/3 since the difference between channel 1 and 3 is 16 dB. This indicates that smaller attenuation difference can result in more stable phase measurement. As for the amplitude, the STD is larger than 1 dB for the channels with weak signal strength. In Set 2, the maximum attenuation difference is 20 dB. It can be found that the phase STD is almost lower than 8° and the amplitude STD is lower than 1 dB. Similarly, the phase STD is lower than 3° for the ports whose difference is smaller than 20 dB. In Set 3, the maximum attenuation difference is 10 dB. It can be observed that the phase STD is lower than 5° in these cases and the amplitude STD is less than 0.5 dB except for the case when the attenuation is larger than 60 dB. From these results, it can be implied that in order to have a stable measurement of amplitude and phase of CSI, the attenuation difference between channels is preferred to be lower than 10 dB. Meanwhile, the attenuation should also be lower than 60 dB.

TABLE V
STANDARD DEVIATION OF AMPLITUDE AND PHASE OF CSI

| Attenuation (dB) | | | Amplitude STD (dB) | | | Phase STD (degree) | | |
|---|---|---|---|---|---|---|---|---|
| Port 1 | Port 2 | Port 3 | Port 1 | Port 2 | Port 3 | Port 2/1 | Port 3/2 | Port 1/3 |
| Set 1 | | | | | | | | |
| 23 | 50 | 50 | 0.16 | 1.31 | 1.81 | 8.77 | 13.35 | 9.91 |
| 26 | 56 | 40 | 0.50 | 1.85 | 0.69 | 14.65 | 15.23 | 2.97 |
| Set 2 | | | | | | | | |
| 40 | 20 | 30 | 1.19 | 0.22 | 0.34 | 7.97 | 2.38 | 8.07 |
| 30 | 40 | 50 | 0.16 | 0.31 | 0.79 | 2.16 | 7.07 | 6.72 |
| 50 | 50 | 30 | 0.90 | 0.72 | 0.20 | 7.49 | 5.06 | 5.53 |
| 56 | 40 | 36 | 1.01 | 0.51 | 0.50 | 5.76 | 1.1 | 5.68 |
| 50 | 60 | 40 | 0.32 | 0.64 | 0.15 | 5.38 | 5.13 | 1.93 |
| Set 3 | | | | | | | | |
| 60 | 56 | 66 | 0.54 | 0.53 | 0.79 | 2.67 | 4.34 | 4.87 |
| 53 | 50 | 63 | 0.20 | 0.21 | 0.48 | 1.69 | 3.56 | 3.40 |
| 23 | 26 | 20 | 0.19 | 0.23 | 0.15 | 1.64 | 1.37 | 1.30 |
| 40 | 43 | 46 | 0.44 | 0.45 | 0.45 | 1.02 | 1.38 | 1.33 |
| 33 | 30 | 36 | 0.20 | 0.20 | 0.24 | 1.48 | 1.62 | 1.76 |
| 50 | 56 | 53 | 0.22 | 0.37 | 0.22 | 2.34 | 2.23 | 1.25 |

*E. Discussion: Active Control for Variation Mitigation*

From the results with extensive experiments, it was implied that the measurement accuracy and stability are jointly affected by the absolute signal strength and the relative strength between channels. Taking both the RSSI and CSI into consideration, it suggests that the signal path loss and the attenuation difference are preferred to be lower than 60 dB and 10 dB, respectively, to obtain stable measurement results. Therefore, when developing the sensing applications, it is recommended to meet these criteria.

Obviously, it is not possible to manipulate the environment during wireless sensing. But there are still actions can be taken to mitigate the measurement variation. As shown in Fig. 12, when the maximum attenuation difference is larger than 30 dB, the measurement of the weak channel is impossible since the readout of CSI were zeros. When the maximum difference is around 30 dB as shown in Set 1 of Table V, although it is possible to measure the amplitude and phase, the amplitude and phase STD is still very large. It should be noted that the attenuation of the channels in Set 1 is less than 60 dB. Based on the findings in this paper, this situation can be improved by adding attenuation to the strongest channel for balancing the signal strength among channels. When the maximum difference is 20 dB as shown in Set 2 of Table V, the amplitude and phase STD are reduced but still larger than the cases in Set 3. Therefore, an active control mechanism can be incorporated into the measurement system. By this mechanism, the signal balance can be adjusted according to the measured signal power levels to mitigate the measurement variation. With stable measurement results, it is promising to design more precise sensing technologies with WiFi signal in the future.

## VI. CONCLUSION

In this paper, the measurement of RSSI and CSI and the variation of the amplitude and phase information with commodity WiFi device was comprehensively investigated. Firstly, the calculation method of RSSI and the amplitude/phase of CSI was put forward in detail. Meanwhile, the relationship between RSSI and CSI was also illustrated. By experiment, it is demonstrated that the path loss of the channels can be correctly represented with the proposed calculation method. Further, the measurement variation of amplitude and phase information was investigated by extensive experiments. It was found that the variation is affected by not only the absolute signal strength but also the signal balance among different channels. If the signal strength is unbalanced, the CSI cannot be measured even with relatively strong signal. From the extensive results, it is revealed that the attenuation of the channel and the difference between channels should not exceed 60 dB and 10 dB, respectively, to obtain stable measurement. The active control mechanism is suggested to mitigate the measurement variation. Based on the findings and proposed criteria of this work, more precise sensing technologies is promising to be developed with reliable measurement results.